\documentclass[superscriptaddress,twocolumn,showpacs,preprintnumbers,amsmath,
amssymb,10pt,aip,jcp]{revtex4}
\usepackage{graphicx}
\usepackage{color}
\begin{document}
\title{From tunneling to contact in a magnetic atom: the non-equilibrium Kondo effect}
\author{Deung-Jang Choi}
\affiliation{CIC nanoGUNE, Tolosa Hiribidea 78, 20018 Donostia-San Sebastian, Spain}
\affiliation{IPCMS, CNRS UMR 7504, Universit\'{e} de Strasbourg, 67034 Strasbourg, France}
\author{Paula Abufager}
\affiliation{Instituto de F\'{i}sica de Rosario, Consejo Nacional de Investigaciones
Cient\'{i}ficas y T\'ecnicas (CONICET) and Universidad Nacional de Rosario,
Av. Pellegrini 250 (2000) Rosario, Argentina}
\author{Laurent Limot}
\affiliation{IPCMS, CNRS UMR 7504, Universit\'{e} de Strasbourg, 67034 Strasbourg, France}
\author{Nicol\'as Lorente}
\affiliation{Centro de F{\'{\i}}sica de Materiales
CFM/MPC (CSIC-UPV/EHU), Paseo Manuel de Lardizabal 5, 20018 Donostia-San Sebasti\'an, Spain}
\affiliation{Donostia International Physics Center (DIPC), Paseo Manuel de Lardizabal 4, 20018 Donostia-San Sebasti\'an, Spain}

\begin{abstract}
A low-temperature scanning tunneling microscope was employed to study the
differential conductance in an atomic junction formed by an adsorbed Co
atom on a Cu(100) surface and a copper-covered tip. A zero-bias anomaly
(ZBA) reveals spin scattering off the Co atom, which is assigned to a
Kondo effect. The ZBA exhibits  a characteristic asymmetric lineshape when
electrons tunnel between tip and sample, while upon the tip-Co contact it
symmetrizes and broadens. Through density functional theory calculations
and the non-equilibrium non-crossing approximation we show that the lineshape
broadening is mainly a consequence of the additional coupling to the
tip, while non-equilibrium effects only modify the large-bias tails of
the ZBA.
\end{abstract}

\date{\today}

\maketitle

\section{Introduction}

The Kondo effect has a long story of research, from the anomalous
low-temperature conductance of noble metals~\cite{Haas}, the explanation
in terms of spin-flip scattering~\cite{Kondo}, to its study in quantum
dots~\cite{Goldhaber,dots}, constrictions~\cite{reyes} and extended
nanosytems~\cite{CNT}. The use of the scanning tunneling microscope (STM)
has undoubtedly revived the research activity around the Kondo effect. The
STM joins unprecedented spatial control of an atomic-scale electrode
with high-resolution in energy. In this way, the low-energy scale
of the Kondo effect can be studied with subatomic precision in single
atoms~\cite{Li_1998,Madhavan_1998,Nagaoka,ternes,Otte_2008,Pruser_2011,Bergmann_2015}
or molecules\cite{molecules,Hla,Ormaza_2016}.

Additionally, STM offers the possibility of creating well-controlled
atomic point contacts by vertically approaching the tip to
an adsorbed atom (adatom hereafter)~\cite{Kroger_2008}. This
drives the microscope in a new regime, where the tip stops
being a non-intrusive probe, to become part of the system to
characterize. The sharp zero-bias anomaly (ZBA) in the differential
conductance, which is the spectroscopic manifestation of the Kondo
effect, is modified by the strong interaction of the adatom with the
tip~\cite{Neel_2007,Choi_2012,Choi_2016}. The resulting new Kondo system
is not just the one of an increased electron bath, but the one in the
presence of a sizeable electron flow. Important theoretical efforts have
been recently given to the study of this so-called non-equilibrium Kondo
regime~\cite{Hershfield_1991,Wingreen_1993,Hettler_1998,Rosch_2001,Monreal_2005,Roermund_2010,Roura_2010,Cohen_2014}.

A simple-minded picture of non-equilibrium effects would consider
two biased Fermi
energies that should lead to the appearance of two Kondo peaks
and to an enhanced and
possibly distorted broadening of the spectral
features.
A more subtle effect is bias-induced decoherence~\cite{Wingreen_1993,Rosch_2001,Monreal_2005,Roermund_2010}.
Indeed, the inequivalent Fermi levels create the situation where
 closed channels for one electrode can be open channels
for the other electrode. Hence, virtual transitions involving the atomic
levels become real transitions. This leads to a lifetime
of intermediate charged states of the atom and thus to the
interruption and hence
decoherence of the spin flips, which ultimately eliminates the Kondo
effect.
Non-equilibrium effects can therefore be expected to profoundly affect the ZBA,
but, to date, little is known on how these effects actually modify the
ZBA line shape.

To study the above effects, we use a
low-temperature STM and form a single-atom contact with a Co atom on
Cu(100). A ZBA is found, which symmetrizes and broadens
as the tip is pushed into contact with the atom. To rationalize this
behavior, we performed density functional theory (DFT) calculations and
transport calculations. The excellent agreement with the measured
conductance over the atom, allows us to make a simple Kondo model using
the non-crossing approximation in non-equilibrium~\cite{Hettler_1998}, and
unravel the contribution of non-equilibrium effects to the experimental
ZBA. We find that in first approximation, the modifications of the ZBA
in the contact regime are mainly driven by the increased coupling to
the electrodes, while the role of the multiple chemical potentials and
non-equilibrium decoherence affect the larger-bias features, i.e. the
tails of the ZBA. This study shows that the Kondo temperatures extracted
in single-atom contacts~\cite{Choi_2012,Choi_2016} reflect the enhanced
interaction of the magnetic impurity with its metallic environment.

\section{Experiment}

\begin{figure}[ht]
\includegraphics[width=1\columnwidth]{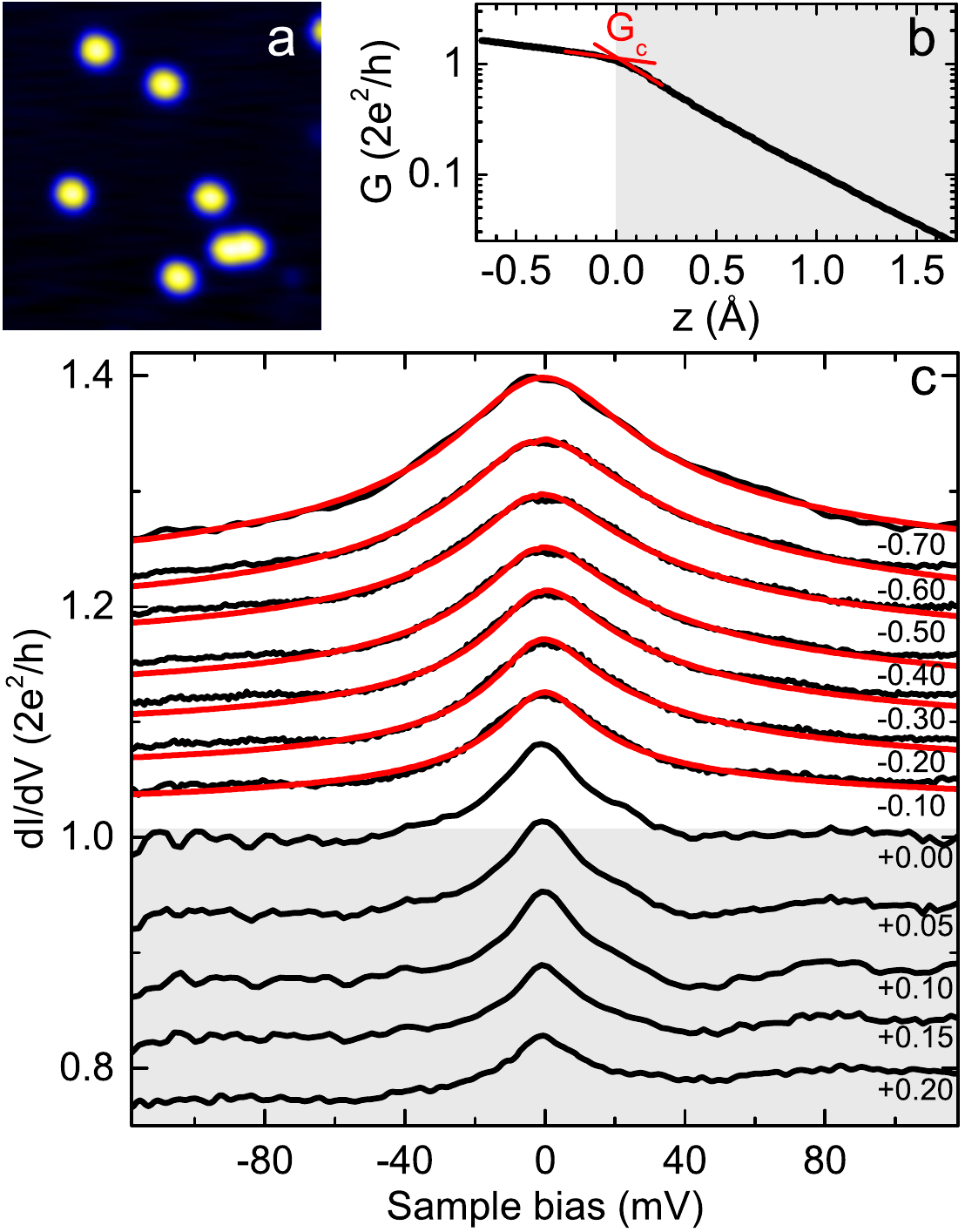}
\caption{\label{stm} {STM measurements on a Co adatom on Cu(100). $(a)$
Topographic image of single Co atoms on Cu(100) (100 pA, 100 mV,
$8\times8$~nm$^2$). $(b)$ Conductance, $G$, versus $z$ curve above a Co
atom; the tunneling regime is indicated by a gray background. The
curves are acquired at a fixed bias of $V=-160$~mV. The contact
conductance, $G_c$, is extracted by approximating the conductance
data in the contact and tunneling regions by straight lines (solid red
lines)~\cite{Neel_2009}. Their point of intersection defines $G_c$. $(c)$
Set of differential conductance spectra acquired with different tip
excursions (indicated in ~{\AA} on the right of the panel); the tip
was verified to have a flat electronic structure in the bias range
presented. Note that a higher tip excursion produces a higher background
in the spectrum. The solid red lines correspond to Frota fits. In the
tunneling regime, which is indicated by a gray background, the ZBA has
a step-like shape. This Fano-like profile results from the interference
between tunneling into the Kondo resonance and tunneling directly into
the substrate~\cite{Li_1998,Madhavan_1998,ternes}.  }
}
\end{figure}

We used an ultra-high vacuum ($<10^{-10}$ mbar) and low temperature (4.4 K)
STM to measure the differential conductance of Co on Cu(100). The Cu(100)
surface, as well as the tungsten tips employed, were cleaned \textit{in
vacuo} by sputter/anneal cycles. The tungsten tips were further prepared
by indentation into the surface to cover their apex with copper. Cobalt
atoms were dosed on the cold Cu(100) substrate by heating a Co wire
(99.99\% purity) resulting in a low concentration of $5\times10^{-3}$
monolayers, Fig.~\ref{stm} $(a)$.

To monitor how the ZBA changes with tip displacement $z$, conductance
versus $z$ curves were first obtained by vertically approaching the
tip from above a target cobalt atom. A typical curve is presented in
Fig.~\ref{stm} $(b)$. A tip displacement of $z=0$ defines the boundary
between the tunneling regime ($z>0$) and the contact regime ($z<0$). Prior
to and after contact the presence of the tip causes readjustments of the
adatom adsorption geometry~\cite{Limot_2005,Ternes_2011} that we will
describe in the theory section. We find that the contact conductance is
$G_c=1.04\pm0.05$ (in units of $2e^2/h$), which demonstrates that the tips
employed have a monoatomically sharp apex~\cite{Neel_2009,Tao_2010}. The
contact geometry has therefore a bottleneck structure comprising a Cu atom
at the tip apex and the Co adatom on Cu(100), Fig.~\ref{scheme}.

The differential conductance as a function of applied bias shows a
characteristic evolution with tip excursion~\cite{Choi_2012,Choi_2016},
Fig.~\ref{stm} $(c)$. The spectra were measured using a lock-in amplifier
(modulation: $500$~$\mu$V~rms, frequency of $712$~Hz) at selected tip
excursions above a Co atom. In the tunneling regime, the ZBA exhibits
a step-like Fano profile~\cite{Li_1998,Madhavan_1998}, while in the
contact regime, the ZBA develops into a peak, symmetrizes and broadens
with decreasing $z$. At contact, the differential conductance, $G$, 
of the ZBA as a function of
the sample bias, $V$,  can be fitted by:

\begin{equation}
G(V)=A+h\,g(V),
\nonumber
\end{equation}
where $h$ and $A$ are constants, and $g$ is the Frota function~\cite{Frota_1986} given by:
\begin{equation}
g(V)=Re \left [\sqrt{\frac{i \Gamma}{i \Gamma+eV-\epsilon_K}}\right].
\label{Frota}
\end{equation}
The center of the peak is located at $\epsilon_K$ and its broadening is controlled by $\Gamma$, related to the Kondo temperature, $T_K$, by~\cite{Pruser_2011}
\[\Gamma= 1.455 \times k_B T_K.\]

Figure~\ref{stm}$(c)$ shows the excellent fits of the ZBA for different
values of $z$ in the contact regime using the above Frota function. We
find a monotonical increase of $T_K$ from 90 K at $z=0$~{\AA} to 200 K
at $z=-0.7$~{\AA}, which reflects a larger screening
of the magnetic impurity as the tip is pressed into contact with the Co
adatom~\cite{Choi_2012,Choi_2016}. However, these Kondo temperatures do not include non-equilibrium
effects that could change the interpretation of the ZBA broadening. This
is the object of the present work.

\section{Theory}

\subsection{Density functional calculations}
\label{dft}

\begin{figure}[ht]
\includegraphics[width=0.5\columnwidth]{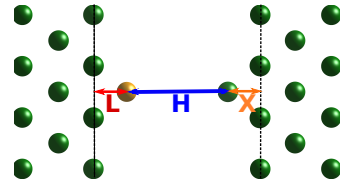}
\caption{\label{scheme}
{
Scheme of the atomic system computed here. Three main
distances are outlined and plotted in Fig.~\ref{relaxation} as
the electrodes are approached.
}
}
\end{figure}

Electronic and geometrical structure optimizations have been perfomed
using  the spin-polarized generalized gradient approximation (GGA-PBE)~\cite{Perdew1996} to
explore the adsorption of Co atoms on the Cu(100) surface.  
To mimic the STM junction, we introduced a second electrode (a Cu (100) surface 
with a Cu adatom, Fig.~\ref{scheme}) that accounts for the tip, and
evaluated the atomic relaxation
with VASP
\cite{Kresse1993a,Kresse1993b,Kresse1996a,Kresse1996b,Kresse1999,Hafner2008}
and {\sc Siesta}~\cite{Soler2002,Artacho2008}. The transport
calculations have been 
performed with
{\sc TranSiesta}~\cite{Brandbyge2002}.

We  used a plane wave basis set and the projected augmented wave (PAW)
method with an energy cut-off of 400~eV.  
The two surfaces representing substrate and tip
were modeled using a slab geometry with a $3 \times 3$
surface unit cell and 5 layers for the surface holding the Co
atom and 6 layers for the tip electrode.  
The valence-electron wavefunctions
were expanded in a basis set of local orbitals in {\sc Siesta}. 
A double-$\zeta$
plus polarization (DZP) basis set was used to describe the Co and surface-atom electrons.
Diffuse orbitals were used to improve the surface electronic
description and
a single-$\zeta$ plus polarization (SZP) basis set for the copper electrodes.  
The use of a DZP basis set to describe the adsorbate states is mandatory
in order to yield correct transmission functions~\cite{Abufager2015}.

The k-point sampling was converged at $7 \times 7$,
although the sampling was $13 \times 13$ for the transmission calculations. 
These transport calculations were carried out from first-principles with a
method based on nonequilibrium Green's functions (NEGF) combined with
DFT as implemented in the {\sc TranSiesta} package~\cite{Brandbyge2002}.

Similar calculations using the single-adsorbate Korringa-Kohn-Rostoker
approach were published by the Mertig group~\cite{Polok_2011}. Our
calculations confirmed theirs with some difference  
associated with the particular DFT implementation.

\begin{figure}[ht]
\includegraphics[width=1.0\columnwidth]{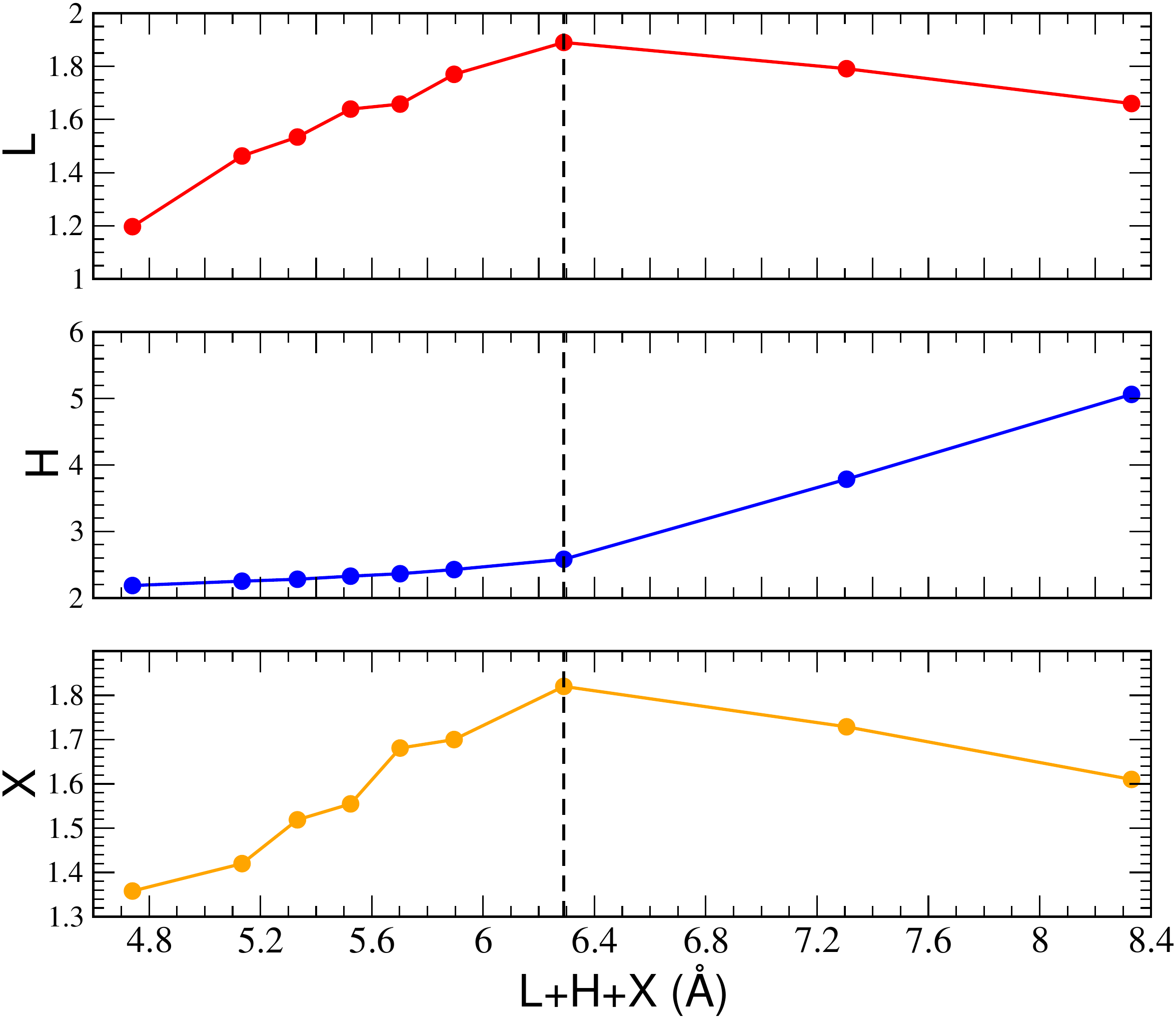}
\caption{\label{relaxation} {Geometrical results of the relaxation
of a Co adatom as a single-atom Cu tip is approached. The vertical
dashed line denotes the point of contact that corresponds to
the formation of an equilibrated covalent bond among all relevant atoms.
Following the scheme of Fig.~\ref{scheme}, the distance of the Co adatom to the Cu(100) substrate is denoted
as $L$. The tip apex to adsorbate distance is $H$ and $X$ is the distance of the apex Cu atom
to the tip's surface. All distances are in \AA. 
After contact, pressing the tip against the adatom does not lead to a significant reduction
of the Co-Cu distance ($H$).
}
}
\end{figure}

The sequential relaxation of the surface structure as the tip approaches
the Co adatom is shown in Fig.~\ref{relaxation}. 
When the tip is far from the substrate, the
relaxation is monotonous and both the Co and Cu adatoms readapt
to the shrinking dimensions of the tunneling junction. As the tip
approaches, the $d$-electrons of both substrates start to hybridize
forming covalent bonds. When the distance of the Co adatom to the
surface is roughly the same as the distance to the Cu tip adatom, there
is a sharp transition. The covalent bonds of the Co adatom are roughly
as strong with the surface as with the tip. This equilibration of the
covalent bonds leads us to define this configuration as the point of contact. Roughly,
all evaluated distances behave similarly in each of the two regions
that the conctact point separates. Before contact, we find that
tunneling is a good description of most of the transport features,
while after contact important changes happen in the way electrons
are transmitted between tip and sample.  Increasing the tip pressure
on the Co adatom leads to pushing the Co adatom to the substrate ($L$
in Fig.~\ref{relaxation}), as well as pushing the Cu adatom to the tip
($X$ in Fig.~\ref{relaxation}). Remarkably, the two-adatom distance
stays basically constant ($H$ in Fig.~\ref{relaxation}).

The detailed electronic structure of the Co adatom changes 
in the contact region. There is a re-ordering of the minority spin $d$-manifold
that affects the transmission properties of the adatom when the tip
starts interacting with it. However, the overall properties slightly
change. The Co adatom can be described as in a 3$d^8$ (S=1)
configuration. This description, corresponding to two Bohr magnetons, is maintained even for the shortest
tip-surface distances we have evaluated.

\begin{figure}[ht]
\includegraphics[width=1.0\columnwidth]{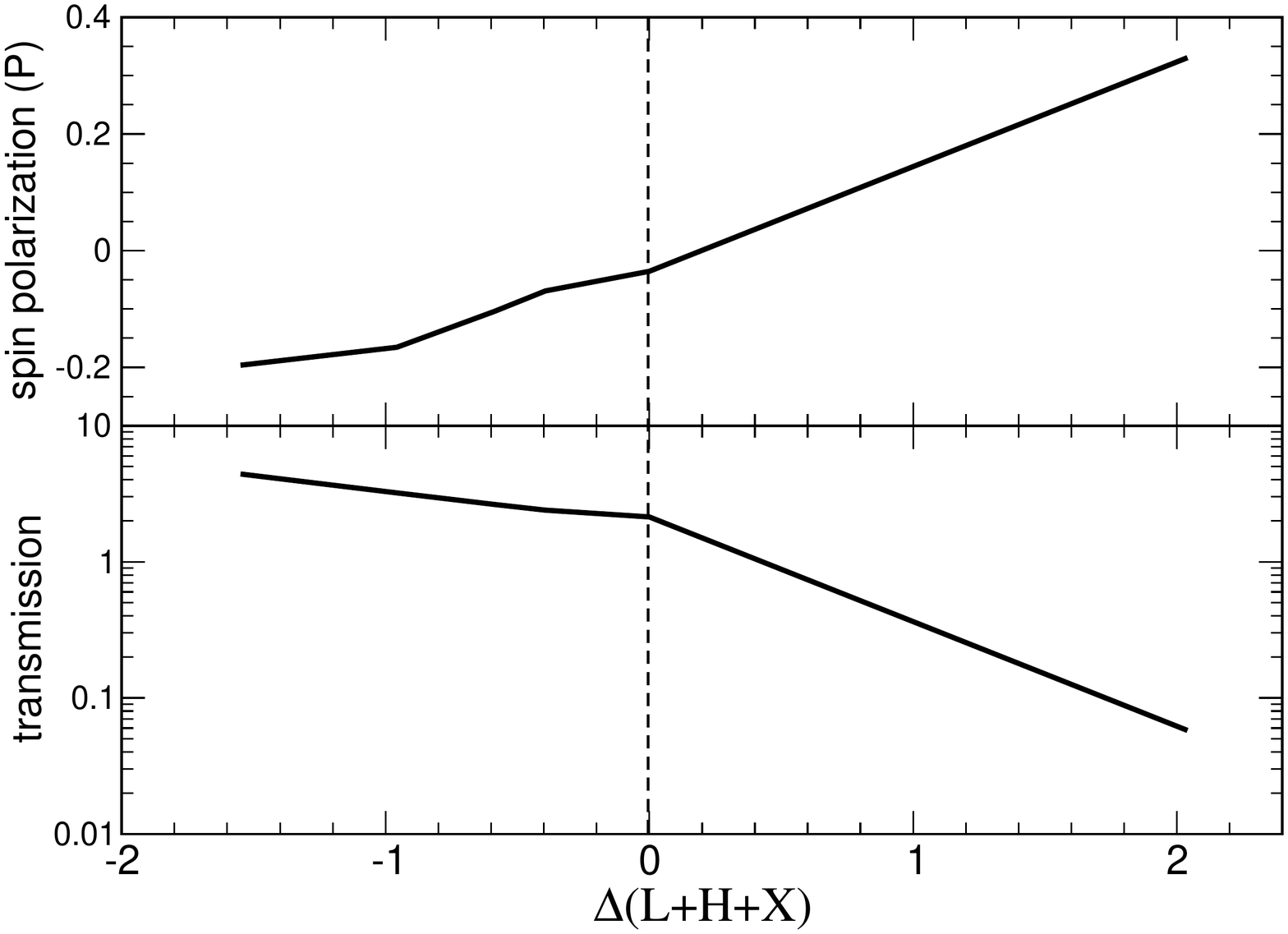}
\caption{\label{transmission} {
Spin polarization and electron transmission at 
the tip-surface distances $\Delta$ (in \AA), measured from the point of contact
that makes it directly comparable with the experimental $z$, Fig.~\ref{stm}. 
The spin polarization, Eq.~(\ref{P}), is defined as the
spin polarization of the electron transmission, and the total
electron transmission is the sum of transmissions for each spin polarization.
Again, contact marks a slope change in the transmission function, stabilizing
the rise of the transmission as the tip is pressed against the substrate.
The spin polarization dramatically changes sign. At contact it is almost zero,
and reverses the polarity as the tip is pressed against the substrate. 
}
}
\end{figure}

In agreement with Ref.~\onlinecite{Polok_2011} we find that transport in the tunneling
region basically takes place through the $sp$ electrons of the Co atom. Figure~\ref{transmission}
shows a rather exponential decay of the transmission prior to contact.
The transmission is due to the majority spin  $sp$ electrons, which leads
to a positive spin polarization  as defined by:
\begin{equation}
P=\frac{T_\uparrow (E_F) - T_\downarrow (E_F)}{T_\uparrow (E_F) + T_\downarrow (E_F)}
\label{P}
\end{equation}
where the transmission per spin $\sigma$, $T_\sigma$, is evaluated at
the Fermi energy, $E_F$.  As the tip approaches, the reorientation and
hybridation of the Co electrons with the tip,  lead to an increase of
the $d$-electron contribution to the transmission. As a consequence the
minority spin $d$ electrons dominate the transmission when the system is
well into the contact region, and the spin polarization of the current
changes sign, Fig.~\ref{transmission}. At contact, the electron current
becomes unpolarized. This effect is purely due to the opening of the
minority spin $d$ channels that compensate the majority spin $sp$
channels, and not to the screening of the Co magnetic moment by the
tip. As we said in the previous paragraph, the absolute value of the
magnetic moment of the Co adatom is basically constant regardless of the
position of the tip.  Finally, as the $d$-electrons start to saturate the
transmission through the Co adatom, the transmission levels off deep
in the contact region, Fig.~\ref{transmission}. For shorter distances
the transmission will continue increasing due to the direct
contribution between Cu atoms.

\begin{figure}[ht]
\includegraphics[width=1.0\columnwidth]{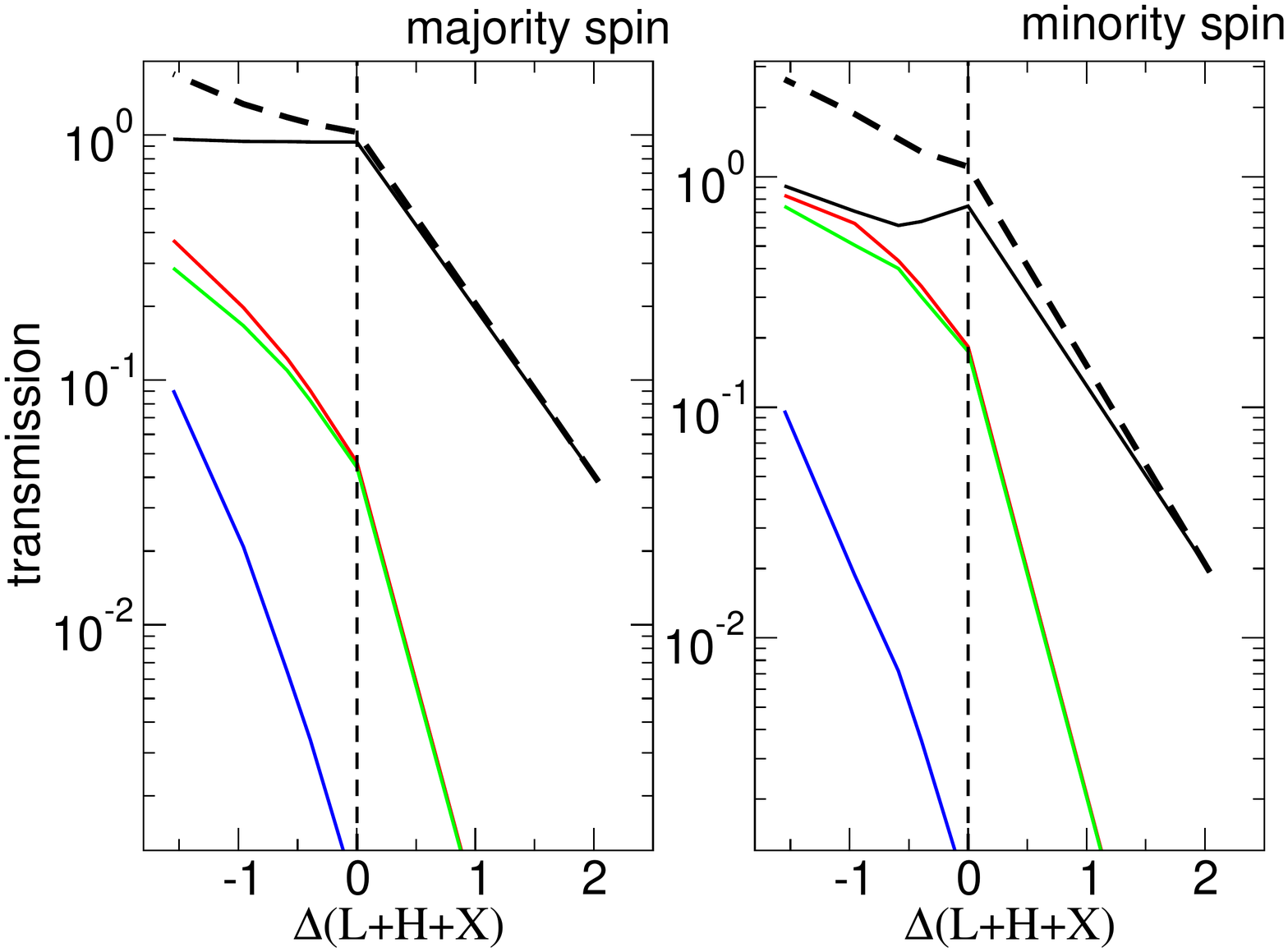}
\caption{\label{Eigenchannels} {
Transmission per spin at the Fermi energy. The thick dashed line is
the total transmission. The other lines correspond to the
transmissions at the Fermi energy of the first four eigenchannel
components. In the tunneling region one eigenchannel controls
the transmission. In contact, the contribution of the $d$ electrons
to transmission leads to the opening of new channels.
The nature of the eigenchannels
change as the tip approaches, from a dominating contribution of the
$sp$ electrons of cobalt in tunneling, to the contribution of $d$ electrons
at contact.
}
}
\end{figure}

The prevailance of minority spins over majority spins in the transmission
is clearly seen in Fig.~\ref{Eigenchannels}. The total transmission
is plotted per spin as well as its eigenchannel composition.  In the
tunneling region, there is a unique eigenchannel that contributes to the
current with a large Co $sp$ component.  As the tip approaches, the $d$
contribution to the tranmission increases, opening  two more eigenchannels
for the minority spin. The majority spin transmission is too small as
can be seen in the right pannel of  Fig.~\ref{Eigenchannels}. In the
contact region, the transmission is formed by three eigenchannels.
The initial one with a large Co $d_{z^2}$ component, where the $sp$
component decreases as the tip approaches the substrate, and two more,
$d_{xz}$ and $d_{yz}$.

The DFT-based transport calculations permit us to conclude that conduction
will take place through the $sp$ electrons of the Co adatom in the
tunneling regime. In contact, the $d_{z^2}$ electrons become dominant
and as the tip pressure increases, contribution from the $d_{xz}$ and
$d_{yz}$ orbitals becomes important.

\subsection{Non-equilibrium Kondo calculations}

Important recent efforts have been devoted to describing the equilibrium
Kondo effect of adsorbed Co on Cu(100). Quantum MonteCarlo (CTQMC)
calculations~\cite{Surer_2012} show the need to describe 
the Kondo effect in this system
as due to fluctuations among three different-charge states ($d^7$, $d^8$ and $d^9$).
They also show
how the final details on the orbital degrees of freedom determine
the Kondo properties of the system. Numerical renormalization
calculations~\cite{Baruselli_2015} performed on a model Anderson
Hamiltonian obtained by fitting its Hartree-Fock solution to DFT
results, show the complexity of the problem and the inadequacy
of DFT to yield quantitative parameters to reproduce the delicate
physics of Kondo impurities.  Nevertheless, these calculations give
robust insight in the orbital electronic structure at the origin of
the Kondo signal of Co on Cu(100). Similar results were obtained
in Refs.~\onlinecite{Jacob_2015,Frank_2015} using the non-crossing
and one-crossing approximations to solve model Anderson Hamiltonians
of Co on Cu(100). Other recent work~\cite{Lounis_2016} using CTQMC on
DFT-fitted Anderson Hamiltonians permits us to rationalize the evolution
of Kondo impurities on different metallic susbtrates.

The above works together with our previous DFT results conclude on
characterizing Co on Cu(100) as a S=1 system, where the $d_{z^2}$
orbital acquires a special meaning due to its enhanced hybridization
both to substrate and tip. Our transport calculations~\ref{dft} however
show that in the tunneling regime, the $sp$-based structure of the Co
adatom dominates the transmission, only when the tip is close enough, does
the $d_{z^2}$ orbital take the lead. This has important consequences
for the differential conductance of the Kondo system.  As shown in
Refs.~\onlinecite{Baruselli_2015,Frank_2015}, Fano profiles develop due
to the interference of transmission paths over the atom. Particularly Frank
and Jacob~\cite{Frank_2015} show that the Co $s$ and $d_{z^2}$ electrons
interfer in ways that can explain the Fano structures experimentally
found in  Co on Cu(100)\cite{Choi_2012}. This picture agrees with our
transport findings. The Fano interference picture stops when
the tip enters in contact with the Co atom.

In section~\ref{dft}, we defined the contact point when the Co-electrode
distance becomes equal to both electrodes. In this case,
the couplings $\Gamma_t$ and $\Gamma_s$ defining the
hybridization of the Co orbitals with both electrodes, tip ($t$) and surface ($s$), 
become equal. Of course these couplings are very different in the tunneling regime.
We evaluate the
differential conductance by computing the numerical derivative of
the Landauer current, $I$, with respect to applied bias, $V$. 
Following the above works~\cite{Baruselli_2015,Jacob_2015,Frank_2015} we take
the conductance to be dominated by the $d_{z^2}$, and hybridization functions  $\Gamma_t$ and $\Gamma_s$
that are mutually proportional. In this case, the many-body
Landauer current becomes~\cite{Meir}:
\begin{equation}
I=\frac{2e}{\hbar} \frac{\Gamma_t \Gamma_s}{\Gamma_t+\Gamma_s} \int [f_t(\omega,V)-f_s(\omega,V)] \rho(\omega,V) d\omega
\label{Landauer}
\end{equation}
where $f_t(\omega,V)$ and $f_s(\omega,V)$ are the equilibrium Fermi distribution
functions for each electrodes, with their respective Fermi energies shifted by $eV$, and $\rho(\omega,V)$
is the non-equilibrium many-body spectral function of the $d_{z^2}$ orbital, also
known as the  non-equilibrium projected density of  states on the $d_{z^2}$ orbital.

To calculate the  non-equilibrium many-body spectral function we resort
to the non-crossing approximation (NCA) where it is particularly
simple to include non-equilibrium effects~\cite{Hettler_1998,Roura_2010}. 
For this, we use our previous formulation of NCA~\cite{Korytar_2011}
extending hybridization functions to include several electrodes
following the equations of Ref.~\onlinecite{Hettler_1998}.
We have assumed a bandwidth of 5 eV where the hybridization function is
completely contained.

\begin{figure}[ht]
\includegraphics[width=1.0\columnwidth]{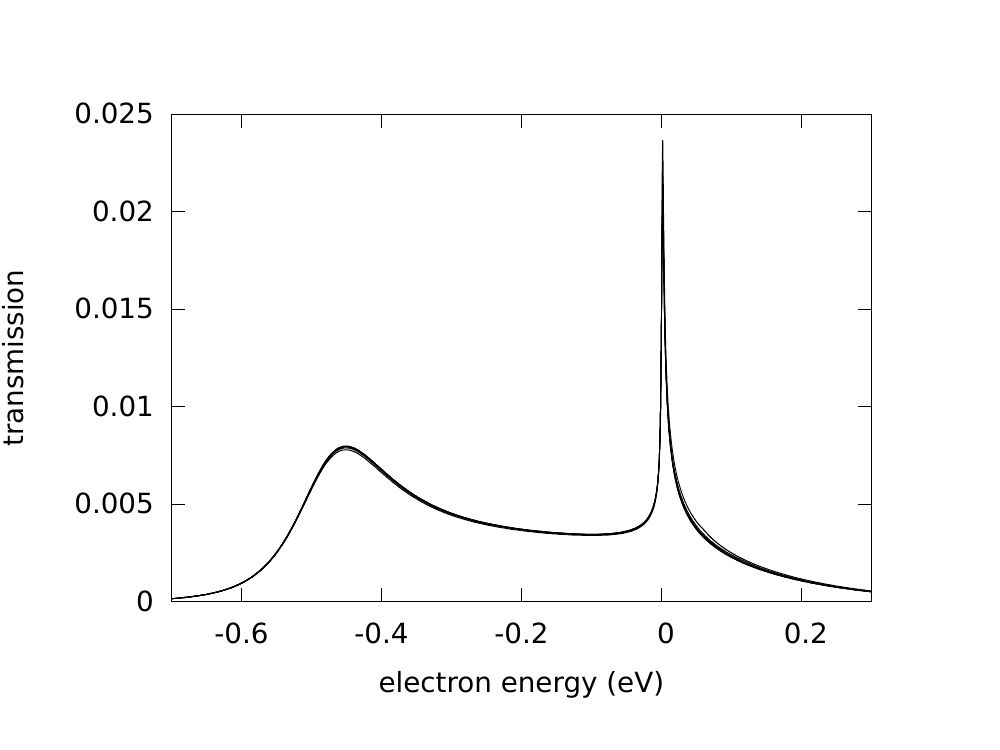}
\caption{\label{tunnel_sp} {Transmission curves as a function
of the electron energy for 10 different biases applied
between tip and substrate between -0.6 V and 0.2 V. The 10 transmission curves
are basically indistinguishable showing no
bias dependence. The coupling of the
tip to the Co atom $d_{z^2}$ is taken to be
30 times smaller than the Co-substrate
coupling. The transmission is directly proportional
to the spectral function in the plotted energy range.}
}
\end{figure}

\begin{figure}[ht]
\includegraphics[width=1.0\columnwidth]{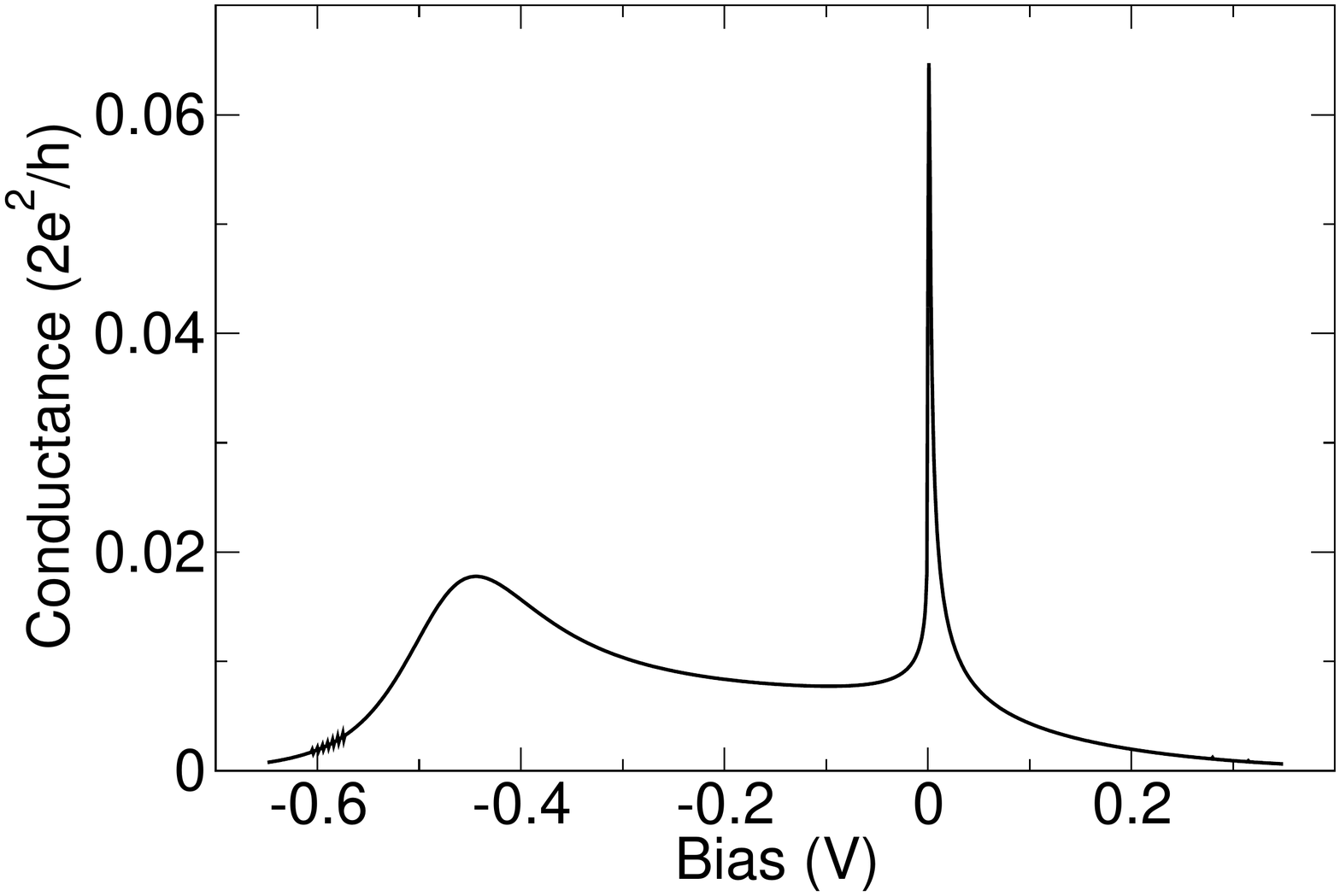}
\caption{\label{tunnel_cond} {
Conductance computed from the previous transmissions for a bias step 
of 1 mV, Fig.~\ref{tunnel_sp}.
The difference is a proportionality constant, showing that the spectral
function determines the differential conductance in the tunneling regime
albeit interference effects between tunneling paths~\cite{Baruselli_2015,Frank_2015}.}
}
\end{figure}

We first take $\Gamma_t=30 \times \Gamma_s$. This corresponds
to the tunneling regime, where the atom is bound to the surface and
follows its density of states as the tip is biased with respect to
the surface.  Our DFT calculations show that transport
takes place through the $sp$ orbitals of the Co atom. To simplify,
we assume that there is a single orbital involved, the $d_{z^2}$,
because we are trying to characterize the evolution of the differential
conductance with bias. For this we simply notice that the adatom spectral
function is constant for all applied biases, see Fig.~\ref{tunnel_sp},
and that the final differential conductance is simply proportional to
the spectral function, Fig.~\ref{tunnel_cond}.  In this situation,
the differential conductance would reflect the ZBA modified by
the intereference among conduction channels giving rise to the Fano
profiles~\cite{Baruselli_2015,Frank_2015}. By fitting a Fano profile,
the values of the Kondo temperature can be estimated as well as other
properties of the Kondo state~\cite{Choi_2012}.

\begin{figure}[ht]
\includegraphics[width=1.0\columnwidth]{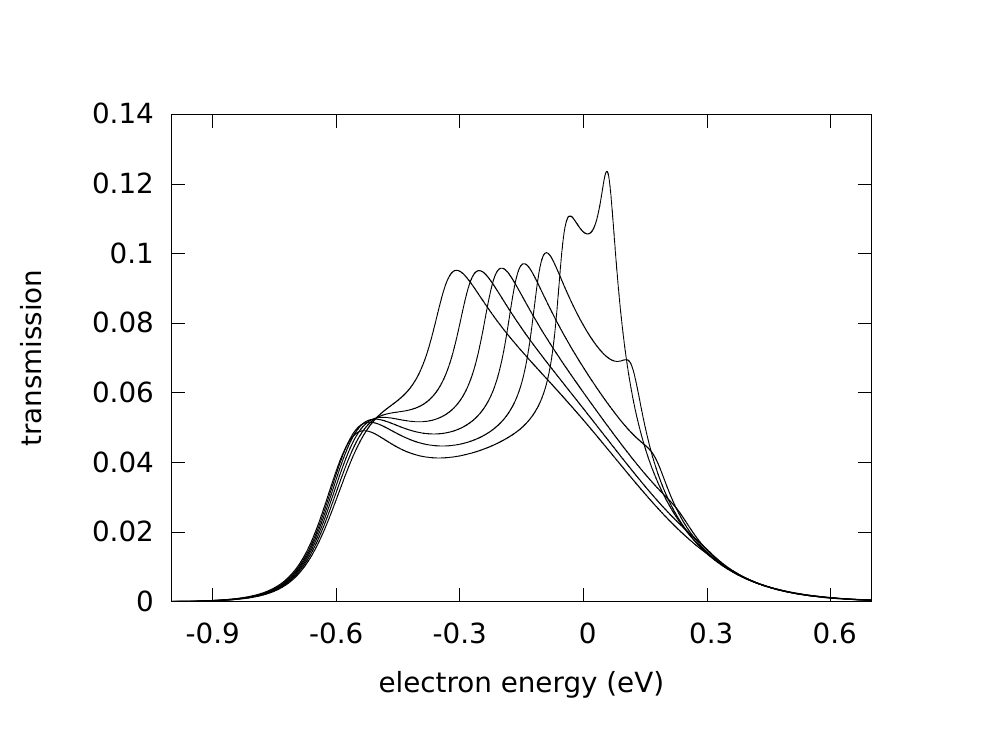}
\caption{\label{contact_sp} {
Transmission curves at contact (equal couplings to both electrodes) for 6 biases
from 0.1 V to 0.6 V. The characteristic splitting of the Kondo
peak is already visible at 0.1 V since the calculations Kondo
temperature is much smaller. Decoherence is also pattent in the
reduction of the peaks and the disappearance of Kondo structure
from the transmission functions.
}
}
\end{figure}

When the tip is pressed on the atom a chemical bond 
forms. 
At contact $\Gamma_t \approx \Gamma_s$.
We use the same couplings to tip and substrate and compute both
spectral functions and differential conductance. Due to the large
coupling to the tip, the electronic structure of the adsorbate
is not fixed to the substrate, see Fig.~\ref{contact_sp}. In
particular, the Kondo peak presents a clear splitting of
the order of the applied bias due to the presence of two Fermi
levels~\cite{Hershfield_1991,Wingreen_1993,Hettler_1998,Rosch_2001, Roura_2010,Cohen_2014}.
As the bias increases, the spectral function grows more
distorted due to the separation of the two Kondo peaks and their
reduction due to the bias-induced decoherence of the scattering
electrons~\cite{Wingreen_1993,Rosch_2001,Monreal_2005,Roermund_2010}.

In the presence of applied bias between electrodes, the Kondo cloud
due to one of the substrates can leak into the other substrate
leading to decoherence of the Kondo ground state. Using the image of
Ref.~\onlinecite{Wingreen_1993} we can envision the Kondo effect
as a coherent spin flip in which an electron coherently hops on and
off the impurity. 
Decoherence takes place if the hopping processes are interrumpted.
This happens
when bias is applied because a new channel opens for
the decay of the electron, when the electron hops on the impurity. 
In other words, the  intermediate state
associated with the electron in the impurity
developes a lifetime that is identified with the bias-induced decoherence
rate~\cite{Wingreen_1993,Rosch_2001,Monreal_2005,Roermund_2010}.  
The decoherence rate~\cite{Rosch_2001} is then the lifetime
of the impurity occupied state (or pseudofermion), given by
twice the imaginary part of the self-energy of the occupied-state
resolvent~\cite{Korytar_2011}.

\begin{figure}[ht]
\includegraphics[width=1.0\columnwidth]{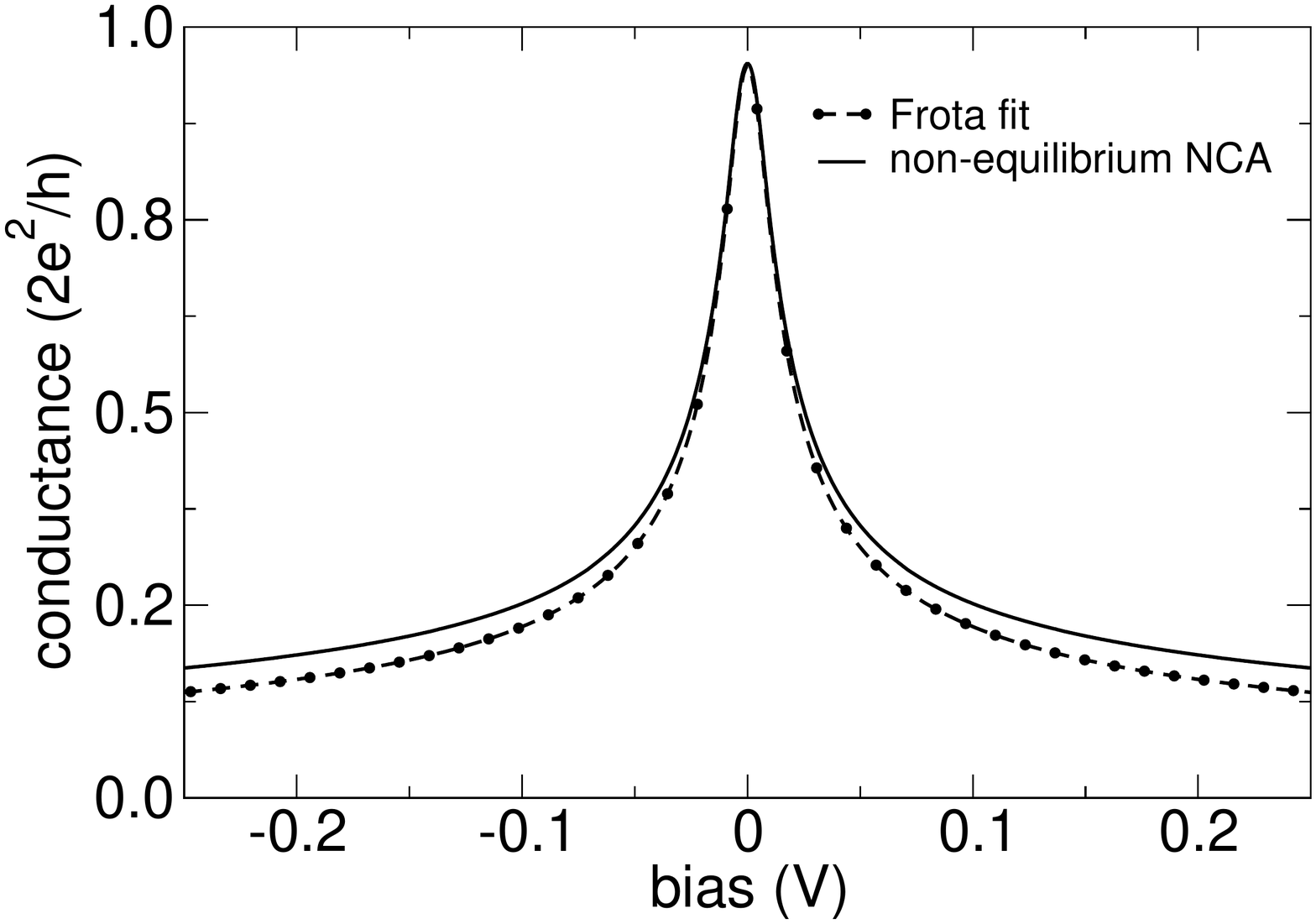}
\caption{\label{contact_cond} {
Conductance computed from a dense grid (steps of 1 mV) of bias-dependent  transmissions, Fig.~\ref{contact_sp}.
The contact regime leads to a broadened
and very symmetric conductance in agreement with the experiment.
The dashed-dotted line is the Frota function fit, Eq.~(\ref{Frota}). The
fit clearly fails to reproduce the larger-bias behavior of the non-equilibrium calculation.
}
}
\end{figure}

The differential conductance at contact does not resemble at all
the spectral function contrary to the above tunneling case. Since
the coupling to the electrodes is symmetric, the conductance has to
be symmetric as found in Fig.~\ref{contact_cond}. This clearly does
not reflect the spectral function, proportional to the transmissions
plotted in Fig.~\ref{contact_sp}. Moreover, the sucessive contributions
of the spectral functions remove any trace of the bias splitting of the
Kondo peak in the density of states. Finally, the conductance features
a broad symmetric peak.

Experimentally, the sharp structure found in tunneling, develops in a
featureless broad peak in excellent agreement with the non-equilibrium
Kondo differential conductance of the above calculations.  Closer
inspection to the calculated spectral functions and conductance
reveals that the width of the conductance peak is roughly twice the
width of the zero-bias spectral function. This factor of two is
due to the $U\rightarrow \infty$ approximation contained in NCA
together with the symmetric condition of
equal coupling to both electrodes,
and is not related to any intrinsic
feature of non-equilibrium Kondo physics. Indeed, the spectral function
of the NCA equilibrium calculation is very asymmetric due to the $U\rightarrow \infty$
requirement of NCA. Hence, simple symmetrization to obtain
a correct differential conductance leads to the factor of two. Only the actual shape of
the peak, as bias increases, reflects the appearance of bias-induced
decoherence and splitting of the Kondo peak. Figure~\ref{contact_cond}
shows a Frota function fit, Eq.~(\ref{Frota}), together
with the calculated non-equilibrium NCA conductance. As expected, at low-bias, the
Frota function is a good fit, while the flatter behavior of NCA conductance at larger bias
signals the appearance of decoherence and non-equilibrium effects. Experimentally then,
most features can be extracted from the equilibrium spectral function
allowing for the correct number of electrodes in the calculation
of the spectral function and of the conductance.  

The  consequence of this study is that the experimentally determined
Kondo temperature does reflect the
Kondo temperature of the Co atom due to the increased coupling to its
enviroment. Non-equilibrium effects are then a small correction to the
estimated temperatures.

\section{Conclusions}

The STM tip can be controllably positioned to create a point
contact with a single magnetic atom between tip and sample
by adjusting the distance between these two non-magnetic electrodes.
This extraordinary setup permits us to study spin-dependent transport
through an atomic constriction with unprecedented accuracy. The
junction atom is a Co atom adsorbed on a Cu(100) surface
and contacted by a Cu-covered STM tip. The differential conductance
at low bias shows spin-flip scattering leading to a ZBA
identified as the Kondo peak. In tunneling, a characteristic
Fano shape reveals that transport takes place through a non-magnetic
channel interfering with the Kondo channel of the
magnetic impurity. In contact the transport and Kondo channels
coalesce and a high-conductance regime leads to a non-equilibrium
Kondo peak.

Using DFT and NEGF transport calculations we can identify
the different channels as well as the magnetic structure
of the atomic junction. Cobalt maintains its magnetic moment
of 2 $\mu_B$ regardless of the tip-substrate
distance, but the transport properties change
as the tip approaches the sample. In tunneling, transport
proceeds by the $sp$ electron structure of the adsorbate.
When the couplings of the atom to tip and substrate equilibrate,
new conduction channels open due to the hybridization of the Co
$d$-electrons to the tip. At contact one single channel dominates
with $d_{z^2}$ character. This orbital possesses the largest
overlap with the holding substrate, making it responsible
for the Kondo signature. Indeed, despite being a S=1 system,
Co on Cu(100) is basically a S=1/2 Kondo system
due to the single $d_{z^2}$ spin~\cite{Choi_2016,Baruselli_2015,Jacob_2015}.

To treat the non-equilibrium Kondo physics when the tip drives
the system into the contact regime, we use the non-equilibrium
non-crossing approximation. In tunneling, the computed conductance
is shown to be proportional to the tunneling spectral function.
This is not true in contact, where the spectral function develops
two Kondo peaks when the applied bias is larger than the
Kondo temperature and the peaks decay by 
bias-induced decoherence. 

Despite this very different regime, the width of the ZBA conductance
is fixed by the equilibrium spectral function, reflecting
the increase of the Kondo temperature due to the larger couplings
to electrodes in the contact regime. Hence, the experimentally
obtained widths are indeed proportional to the Kondo
temperature of the system as the point contact increases
its atomic coordination. 

\noindent\textbf{Acknowledgements}\\

PA and NL sincerely thank Jaime Ferrer, Pablo Rivero and Salva Barraza for
providing us with their cobalt pseudopotential.  DJC acknowledges the
European Union for support under the H2020-MSCA-IF-2014 Marie-Curie
Individual Fellowship programme proposal number 654469. PA
acknowledges the CCT-Rosario Computational Center. Financial
support from CONICET, the Agence Nationale de la Recherche (Grant
No. ANR-13-BS10-0016, ANR-11-LABX-0058 NIE, ANR-10-LABX-0026 CSC) and
MINECO (Grant No. MAT2015-66888-C3-2-R) are gratefully acknowledged.

\bibliography{kondo} 

\begin{thebibliography}{52}
\expandafter\ifx\csname natexlab\endcsname\relax\def\natexlab#1{#1}\fi
\expandafter\ifx\csname bibnamefont\endcsname\relax
  \def\bibnamefont#1{#1}\fi
\expandafter\ifx\csname bibfnamefont\endcsname\relax
  \def\bibfnamefont#1{#1}\fi
\expandafter\ifx\csname citenamefont\endcsname\relax
  \def\citenamefont#1{#1}\fi
\expandafter\ifx\csname url\endcsname\relax
  \def\url#1{\texttt{#1}}\fi
\expandafter\ifx\csname urlprefix\endcsname\relax\def\urlprefix{URL }\fi
\providecommand{\bibinfo}[2]{#2}
\providecommand{\eprint}[2][]{\url{#2}}

\bibitem[{\citenamefont{Haas and Berg}(1936)}]{Haas}
\bibinfo{author}{\bibfnamefont{W.~D.} \bibnamefont{Haas}} \bibnamefont{and}
  \bibinfo{author}{\bibfnamefont{G.~V.~D.} \bibnamefont{Berg}},
  \bibinfo{journal}{Physica} \textbf{\bibinfo{volume}{3}}, \bibinfo{pages}{440
  } (\bibinfo{year}{1936}).

\bibitem[{\citenamefont{Kondo}(1964)}]{Kondo}
\bibinfo{author}{\bibfnamefont{J.}~\bibnamefont{Kondo}},
  \bibinfo{journal}{Progress of Theoretical Physics}
  \textbf{\bibinfo{volume}{32}}, \bibinfo{pages}{37} (\bibinfo{year}{1964}),
  \urlprefix\url{http://ptp.ipap.jp/link?PTP/32/37/}.

\bibitem[{\citenamefont{Goldhaber-Gordon
  et~al.}(1998)\citenamefont{Goldhaber-Gordon, Shtrikman, Mahaly,
  Abusch-Magder, Meirav, and Kastner}}]{Goldhaber}
\bibinfo{author}{\bibfnamefont{D.}~\bibnamefont{Goldhaber-Gordon}},
  \bibinfo{author}{\bibfnamefont{H.}~\bibnamefont{Shtrikman}},
  \bibinfo{author}{\bibfnamefont{D.}~\bibnamefont{Mahaly}},
  \bibinfo{author}{\bibfnamefont{D.}~\bibnamefont{Abusch-Magder}},
  \bibinfo{author}{\bibfnamefont{U.}~\bibnamefont{Meirav}}, \bibnamefont{and}
  \bibinfo{author}{\bibfnamefont{M.~A.} \bibnamefont{Kastner}},
  \bibinfo{journal}{Nature} \textbf{\bibinfo{volume}{391}},
  \bibinfo{pages}{156} (\bibinfo{year}{1998}).

\bibitem[{\citenamefont{Cronenwett et~al.}(1998)\citenamefont{Cronenwett,
  Oosterkamp, and Kouwenhoven}}]{dots}
\bibinfo{author}{\bibfnamefont{S.~M.} \bibnamefont{Cronenwett}},
  \bibinfo{author}{\bibfnamefont{T.~H.} \bibnamefont{Oosterkamp}},
  \bibnamefont{and} \bibinfo{author}{\bibfnamefont{L.~P.}
  \bibnamefont{Kouwenhoven}}, \bibinfo{journal}{Science}
  \textbf{\bibinfo{volume}{281}}, \bibinfo{pages}{540} (\bibinfo{year}{1998}),
  ISSN \bibinfo{issn}{0036-8075},
  \urlprefix\url{http://science.sciencemag.org/content/281/5376/540}.

\bibitem[{\citenamefont{Reyes-Calvo et~al.}(2009)\citenamefont{Reyes-Calvo,
  Fern\'andez-Rossier, Palacios, Jacob, Natelson, and Untiedt}}]{reyes}
\bibinfo{author}{\bibfnamefont{M.}~\bibnamefont{Reyes-Calvo}},
  \bibinfo{author}{\bibfnamefont{J.}~\bibnamefont{Fern\'andez-Rossier}},
  \bibinfo{author}{\bibfnamefont{J.~J.} \bibnamefont{Palacios}},
  \bibinfo{author}{\bibfnamefont{D.}~\bibnamefont{Jacob}},
  \bibinfo{author}{\bibfnamefont{D.}~\bibnamefont{Natelson}}, \bibnamefont{and}
  \bibinfo{author}{\bibfnamefont{C.}~\bibnamefont{Untiedt}},
  \bibinfo{journal}{Nature} \textbf{\bibinfo{volume}{458}},
  \bibinfo{pages}{1150} (\bibinfo{year}{2009}).

\bibitem[{\citenamefont{Nygard et~al.}(2000)\citenamefont{Nygard, Cobden, and
  Lindelof}}]{CNT}
\bibinfo{author}{\bibfnamefont{J.}~\bibnamefont{Nygard}},
  \bibinfo{author}{\bibfnamefont{D.~H.} \bibnamefont{Cobden}},
  \bibnamefont{and} \bibinfo{author}{\bibfnamefont{P.~E.}
  \bibnamefont{Lindelof}}, \bibinfo{journal}{Nature}
  \textbf{\bibinfo{volume}{408}}, \bibinfo{pages}{342} (\bibinfo{year}{2000}).

\bibitem[{\citenamefont{Li et~al.}(1998)\citenamefont{Li, Schneider, Berndt,
  and Delley}}]{Li_1998}
\bibinfo{author}{\bibfnamefont{J.}~\bibnamefont{Li}},
  \bibinfo{author}{\bibfnamefont{W.-D.} \bibnamefont{Schneider}},
  \bibinfo{author}{\bibfnamefont{R.}~\bibnamefont{Berndt}}, \bibnamefont{and}
  \bibinfo{author}{\bibfnamefont{B.}~\bibnamefont{Delley}},
  \bibinfo{journal}{Phys. Rev. Lett.} \textbf{\bibinfo{volume}{80}},
  \bibinfo{pages}{2893} (\bibinfo{year}{1998}),
  \urlprefix\url{http://link.aps.org/doi/10.1103/PhysRevLett.80.2893}.

\bibitem[{\citenamefont{Madhavan et~al.}(1998)\citenamefont{Madhavan, Chen,
  Jamneala, Crommie, and Wingreen}}]{Madhavan_1998}
\bibinfo{author}{\bibfnamefont{V.}~\bibnamefont{Madhavan}},
  \bibinfo{author}{\bibfnamefont{W.}~\bibnamefont{Chen}},
  \bibinfo{author}{\bibfnamefont{T.}~\bibnamefont{Jamneala}},
  \bibinfo{author}{\bibfnamefont{M.~F.} \bibnamefont{Crommie}},
  \bibnamefont{and} \bibinfo{author}{\bibfnamefont{N.~S.}
  \bibnamefont{Wingreen}}, \bibinfo{journal}{Science}
  \textbf{\bibinfo{volume}{280}}, \bibinfo{pages}{567} (\bibinfo{year}{1998}),
  ISSN \bibinfo{issn}{0036-8075},
  \urlprefix\url{http://science.sciencemag.org/content/280/5363/567}.

\bibitem[{\citenamefont{Nagaoka et~al.}(2002)\citenamefont{Nagaoka, Jamneala,
  Grobis, and Crommie}}]{Nagaoka}
\bibinfo{author}{\bibfnamefont{K.}~\bibnamefont{Nagaoka}},
  \bibinfo{author}{\bibfnamefont{T.}~\bibnamefont{Jamneala}},
  \bibinfo{author}{\bibfnamefont{M.}~\bibnamefont{Grobis}}, \bibnamefont{and}
  \bibinfo{author}{\bibfnamefont{M.~F.} \bibnamefont{Crommie}},
  \bibinfo{journal}{Phys. Rev. Lett.} \textbf{\bibinfo{volume}{88}},
  \bibinfo{pages}{077205} (\bibinfo{year}{2002}),
  \urlprefix\url{http://link.aps.org/doi/10.1103/PhysRevLett.88.077205}.

\bibitem[{\citenamefont{Ternes et~al.}(2009)\citenamefont{Ternes, Heinrich, and
  Schneider}}]{ternes}
\bibinfo{author}{\bibfnamefont{M.}~\bibnamefont{Ternes}},
  \bibinfo{author}{\bibfnamefont{A.~J.} \bibnamefont{Heinrich}},
  \bibnamefont{and} \bibinfo{author}{\bibfnamefont{W.-D.}
  \bibnamefont{Schneider}}, \bibinfo{journal}{Journal of Physics: Condensed
  Matter} \textbf{\bibinfo{volume}{21}}, \bibinfo{pages}{053001}
  (\bibinfo{year}{2009}),
  \urlprefix\url{http://stacks.iop.org/0953-8984/21/i=5/a=053001}.

\bibitem[{\citenamefont{Otte et~al.}(2008)\citenamefont{Otte, Ternes, von
  Bergmann, Loth, Brune, Lutz, Hirjibehedin, and Heinrich}}]{Otte_2008}
\bibinfo{author}{\bibfnamefont{A.~F.} \bibnamefont{Otte}},
  \bibinfo{author}{\bibfnamefont{M.}~\bibnamefont{Ternes}},
  \bibinfo{author}{\bibfnamefont{K.}~\bibnamefont{von Bergmann}},
  \bibinfo{author}{\bibfnamefont{S.}~\bibnamefont{Loth}},
  \bibinfo{author}{\bibfnamefont{H.}~\bibnamefont{Brune}},
  \bibinfo{author}{\bibfnamefont{C.~P.} \bibnamefont{Lutz}},
  \bibinfo{author}{\bibfnamefont{C.~F.} \bibnamefont{Hirjibehedin}},
  \bibnamefont{and} \bibinfo{author}{\bibfnamefont{A.~J.}
  \bibnamefont{Heinrich}}, \bibinfo{journal}{Nat. Phys.}
  \textbf{\bibinfo{volume}{4}}, \bibinfo{pages}{847} (\bibinfo{year}{2008}),
  \urlprefix\url{http://dx.doi.org/10.1038/nphys1072}.

\bibitem[{\citenamefont{Pruser et~al.}(2011)\citenamefont{Pruser, Wenderoth,
  Dargel, Weismann, Peters, Pruschke, and Ulbrich}}]{Pruser_2011}
\bibinfo{author}{\bibfnamefont{H.}~\bibnamefont{Pruser}},
  \bibinfo{author}{\bibfnamefont{M.}~\bibnamefont{Wenderoth}},
  \bibinfo{author}{\bibfnamefont{P.~E.} \bibnamefont{Dargel}},
  \bibinfo{author}{\bibfnamefont{A.}~\bibnamefont{Weismann}},
  \bibinfo{author}{\bibfnamefont{R.}~\bibnamefont{Peters}},
  \bibinfo{author}{\bibfnamefont{T.}~\bibnamefont{Pruschke}}, \bibnamefont{and}
  \bibinfo{author}{\bibfnamefont{R.~G.} \bibnamefont{Ulbrich}},
  \bibinfo{journal}{Nat. Phys.} \textbf{\bibinfo{volume}{7}},
  \bibinfo{pages}{203} (\bibinfo{year}{2011}).

\bibitem[{\citenamefont{von Bergmann et~al.}(2015)\citenamefont{von Bergmann,
  Ternes, Loth, Lutz, and Heinrich}}]{Bergmann_2015}
\bibinfo{author}{\bibfnamefont{K.}~\bibnamefont{von Bergmann}},
  \bibinfo{author}{\bibfnamefont{M.}~\bibnamefont{Ternes}},
  \bibinfo{author}{\bibfnamefont{S.}~\bibnamefont{Loth}},
  \bibinfo{author}{\bibfnamefont{C.~P.} \bibnamefont{Lutz}}, \bibnamefont{and}
  \bibinfo{author}{\bibfnamefont{A.~J.} \bibnamefont{Heinrich}},
  \bibinfo{journal}{Phys. Rev. Lett.} \textbf{\bibinfo{volume}{114}},
  \bibinfo{pages}{076601} (\bibinfo{year}{2015}),
  \urlprefix\url{http://link.aps.org/doi/10.1103/PhysRevLett.114.076601}.

\bibitem[{\citenamefont{Park et~al.}(2000)\citenamefont{Park, Pasupathy,
  Goldsmith, Chang, Yaish, Petta, Rinkoski, Sethna, Abru\~na, McEuen
  et~al.}}]{molecules}
\bibinfo{author}{\bibfnamefont{J.}~\bibnamefont{Park}},
  \bibinfo{author}{\bibfnamefont{A.~N.} \bibnamefont{Pasupathy}},
  \bibinfo{author}{\bibfnamefont{J.~I.} \bibnamefont{Goldsmith}},
  \bibinfo{author}{\bibfnamefont{C.}~\bibnamefont{Chang}},
  \bibinfo{author}{\bibfnamefont{Y.}~\bibnamefont{Yaish}},
  \bibinfo{author}{\bibfnamefont{J.~R.} \bibnamefont{Petta}},
  \bibinfo{author}{\bibfnamefont{M.}~\bibnamefont{Rinkoski}},
  \bibinfo{author}{\bibfnamefont{J.~P.} \bibnamefont{Sethna}},
  \bibinfo{author}{\bibfnamefont{H.~D.} \bibnamefont{Abru\~na}},
  \bibinfo{author}{\bibfnamefont{P.~L.} \bibnamefont{McEuen}},
  \bibnamefont{et~al.}, \bibinfo{journal}{Nature}
  \textbf{\bibinfo{volume}{408}}, \bibinfo{pages}{342} (\bibinfo{year}{2000}).

\bibitem[{\citenamefont{Perera et~al.}(2010)\citenamefont{Perera, Kulik, Iancu,
  Dias~da Silva, Ulloa, Marzari, and Hla}}]{Hla}
\bibinfo{author}{\bibfnamefont{U.~G.~E.} \bibnamefont{Perera}},
  \bibinfo{author}{\bibfnamefont{H.~J.} \bibnamefont{Kulik}},
  \bibinfo{author}{\bibfnamefont{V.}~\bibnamefont{Iancu}},
  \bibinfo{author}{\bibfnamefont{L.~G. G.~V.} \bibnamefont{Dias~da Silva}},
  \bibinfo{author}{\bibfnamefont{S.~E.} \bibnamefont{Ulloa}},
  \bibinfo{author}{\bibfnamefont{N.}~\bibnamefont{Marzari}}, \bibnamefont{and}
  \bibinfo{author}{\bibfnamefont{S.-W.} \bibnamefont{Hla}},
  \bibinfo{journal}{Physical Review Letters} \textbf{\bibinfo{volume}{105}},
  \bibinfo{pages}{106601} (\bibinfo{year}{2010}),
  \urlprefix\url{http://link.aps.org/doi/10.1103/PhysRevLett.105.106601}.

\bibitem[{\citenamefont{Ormaza et~al.}(2016)\citenamefont{Ormaza, Robles,
  Bachellier, Abufager, Lorente, and Limot}}]{Ormaza_2016}
\bibinfo{author}{\bibfnamefont{M.}~\bibnamefont{Ormaza}},
  \bibinfo{author}{\bibfnamefont{R.}~\bibnamefont{Robles}},
  \bibinfo{author}{\bibfnamefont{N.}~\bibnamefont{Bachellier}},
  \bibinfo{author}{\bibfnamefont{P.}~\bibnamefont{Abufager}},
  \bibinfo{author}{\bibfnamefont{N.}~\bibnamefont{Lorente}}, \bibnamefont{and}
  \bibinfo{author}{\bibfnamefont{L.}~\bibnamefont{Limot}},
  \bibinfo{journal}{Nano Lett.} \textbf{\bibinfo{volume}{16}},
  \bibinfo{pages}{588} (\bibinfo{year}{2016}).

\bibitem[{\citenamefont{Kr{\"o}ger et~al.}(2008)\citenamefont{Kr{\"o}ger,
  N{\'e}el, and Limot}}]{Kroger_2008}
\bibinfo{author}{\bibfnamefont{J.}~\bibnamefont{Kr{\"o}ger}},
  \bibinfo{author}{\bibfnamefont{N.}~\bibnamefont{N{\'e}el}}, \bibnamefont{and}
  \bibinfo{author}{\bibfnamefont{L.}~\bibnamefont{Limot}},
  \bibinfo{journal}{Journal of Physics: Condensed Matter}
  \textbf{\bibinfo{volume}{20}}, \bibinfo{pages}{223001}
  (\bibinfo{year}{2008}),
  \urlprefix\url{http://stacks.iop.org/0953-8984/20/i=22/a=223001}.

\bibitem[{\citenamefont{N\'eel et~al.}(2007)\citenamefont{N\'eel, Kr\"oger,
  Limot, Palotas, Hofer, and Berndt}}]{Neel_2007}
\bibinfo{author}{\bibfnamefont{N.}~\bibnamefont{N\'eel}},
  \bibinfo{author}{\bibfnamefont{J.}~\bibnamefont{Kr\"oger}},
  \bibinfo{author}{\bibfnamefont{L.}~\bibnamefont{Limot}},
  \bibinfo{author}{\bibfnamefont{K.}~\bibnamefont{Palotas}},
  \bibinfo{author}{\bibfnamefont{W.~A.} \bibnamefont{Hofer}}, \bibnamefont{and}
  \bibinfo{author}{\bibfnamefont{R.}~\bibnamefont{Berndt}},
  \bibinfo{journal}{Phys. Rev. Lett.} \textbf{\bibinfo{volume}{98}},
  \bibinfo{pages}{016801} (\bibinfo{year}{2007}).

\bibitem[{\citenamefont{Choi et~al.}(2012)\citenamefont{Choi, Rastei, Simon,
  and Limot}}]{Choi_2012}
\bibinfo{author}{\bibfnamefont{D.-J.} \bibnamefont{Choi}},
  \bibinfo{author}{\bibfnamefont{M.~V.} \bibnamefont{Rastei}},
  \bibinfo{author}{\bibfnamefont{P.}~\bibnamefont{Simon}}, \bibnamefont{and}
  \bibinfo{author}{\bibfnamefont{L.}~\bibnamefont{Limot}},
  \bibinfo{journal}{Phys. Rev. Lett.} \textbf{\bibinfo{volume}{108}},
  \bibinfo{pages}{266803} (\bibinfo{year}{2012}),
  \urlprefix\url{http://link.aps.org/doi/10.1103/PhysRevLett.108.266803}.

\bibitem[{\citenamefont{Choi et~al.}(0)\citenamefont{Choi, Guissart, Ormaza,
  Bachellier, Bengone, Simon, and Limot}}]{Choi_2016}
\bibinfo{author}{\bibfnamefont{D.-J.} \bibnamefont{Choi}},
  \bibinfo{author}{\bibfnamefont{S.}~\bibnamefont{Guissart}},
  \bibinfo{author}{\bibfnamefont{M.}~\bibnamefont{Ormaza}},
  \bibinfo{author}{\bibfnamefont{N.}~\bibnamefont{Bachellier}},
  \bibinfo{author}{\bibfnamefont{O.}~\bibnamefont{Bengone}},
  \bibinfo{author}{\bibfnamefont{P.}~\bibnamefont{Simon}}, \bibnamefont{and}
  \bibinfo{author}{\bibfnamefont{L.}~\bibnamefont{Limot}},
  \bibinfo{journal}{Nano Lett.} \textbf{\bibinfo{volume}{0}},
  \bibinfo{pages}{0} (\bibinfo{year}{0}),
  \urlprefix\url{http://dx.doi.org/10.1021/acs.nanolett.6b02617}.

\bibitem[{\citenamefont{Hershfield et~al.}(1991)\citenamefont{Hershfield,
  Davies, and Wilkins}}]{Hershfield_1991}
\bibinfo{author}{\bibfnamefont{S.}~\bibnamefont{Hershfield}},
  \bibinfo{author}{\bibfnamefont{J.~H.} \bibnamefont{Davies}},
  \bibnamefont{and} \bibinfo{author}{\bibfnamefont{J.~W.}
  \bibnamefont{Wilkins}}, \bibinfo{journal}{Phys. Rev. Lett.}
  \textbf{\bibinfo{volume}{67}}, \bibinfo{pages}{3720} (\bibinfo{year}{1991}),
  \urlprefix\url{http://link.aps.org/doi/10.1103/PhysRevLett.67.3720}.

\bibitem[{\citenamefont{Meir et~al.}(1993)\citenamefont{Meir, Wingreen, and
  Lee}}]{Wingreen_1993}
\bibinfo{author}{\bibfnamefont{Y.}~\bibnamefont{Meir}},
  \bibinfo{author}{\bibfnamefont{N.~S.} \bibnamefont{Wingreen}},
  \bibnamefont{and} \bibinfo{author}{\bibfnamefont{P.~A.} \bibnamefont{Lee}},
  \bibinfo{journal}{Phys. Rev. Lett.} \textbf{\bibinfo{volume}{70}},
  \bibinfo{pages}{2601} (\bibinfo{year}{1993}),
  \urlprefix\url{http://link.aps.org/doi/10.1103/PhysRevLett.70.2601}.

\bibitem[{\citenamefont{Hettler et~al.}(1998)\citenamefont{Hettler, Kroha, and
  Hershfield}}]{Hettler_1998}
\bibinfo{author}{\bibfnamefont{M.~H.} \bibnamefont{Hettler}},
  \bibinfo{author}{\bibfnamefont{J.}~\bibnamefont{Kroha}}, \bibnamefont{and}
  \bibinfo{author}{\bibfnamefont{S.}~\bibnamefont{Hershfield}},
  \bibinfo{journal}{Phys. Rev. B} \textbf{\bibinfo{volume}{58}},
  \bibinfo{pages}{5649} (\bibinfo{year}{1998}),
  \urlprefix\url{http://link.aps.org/doi/10.1103/PhysRevB.58.5649}.

\bibitem[{\citenamefont{Rosch et~al.}(2001)\citenamefont{Rosch, Kroha, and
  W\"olfle}}]{Rosch_2001}
\bibinfo{author}{\bibfnamefont{A.}~\bibnamefont{Rosch}},
  \bibinfo{author}{\bibfnamefont{J.}~\bibnamefont{Kroha}}, \bibnamefont{and}
  \bibinfo{author}{\bibfnamefont{P.}~\bibnamefont{W\"olfle}},
  \bibinfo{journal}{Phys. Rev. Lett.} \textbf{\bibinfo{volume}{87}},
  \bibinfo{pages}{156802} (\bibinfo{year}{2001}),
  \urlprefix\url{http://link.aps.org/doi/10.1103/PhysRevLett.87.156802}.

\bibitem[{\citenamefont{Monreal and Flores}(2005)}]{Monreal_2005}
\bibinfo{author}{\bibfnamefont{R.~C.} \bibnamefont{Monreal}} \bibnamefont{and}
  \bibinfo{author}{\bibfnamefont{F.}~\bibnamefont{Flores}},
  \bibinfo{journal}{Phys. Rev. B} \textbf{\bibinfo{volume}{72}},
  \bibinfo{pages}{195105} (\bibinfo{year}{2005}),
  \urlprefix\url{http://link.aps.org/doi/10.1103/PhysRevB.72.195105}.

\bibitem[{\citenamefont{Van~Roermund et~al.}(2010)\citenamefont{Van~Roermund,
  Shiau, and Lavagna}}]{Roermund_2010}
\bibinfo{author}{\bibfnamefont{R.}~\bibnamefont{Van~Roermund}},
  \bibinfo{author}{\bibfnamefont{S.-y.} \bibnamefont{Shiau}}, \bibnamefont{and}
  \bibinfo{author}{\bibfnamefont{M.}~\bibnamefont{Lavagna}},
  \bibinfo{journal}{Phys. Rev. B} \textbf{\bibinfo{volume}{81}},
  \bibinfo{pages}{165115} (\bibinfo{year}{2010}),
  \urlprefix\url{http://link.aps.org/doi/10.1103/PhysRevB.81.165115}.

\bibitem[{\citenamefont{Bas and Aligia}(2010)}]{Roura_2010}
\bibinfo{author}{\bibfnamefont{P.~R.} \bibnamefont{Bas}} \bibnamefont{and}
  \bibinfo{author}{\bibfnamefont{A.~A.} \bibnamefont{Aligia}},
  \bibinfo{journal}{Journal of Physics: Condensed Matter}
  \textbf{\bibinfo{volume}{22}}, \bibinfo{pages}{025602}
  (\bibinfo{year}{2010}),
  \urlprefix\url{http://stacks.iop.org/0953-8984/22/i=2/a=025602}.

\bibitem[{\citenamefont{Cohen et~al.}(2014)\citenamefont{Cohen, Gull, Reichman,
  and Millis}}]{Cohen_2014}
\bibinfo{author}{\bibfnamefont{G.}~\bibnamefont{Cohen}},
  \bibinfo{author}{\bibfnamefont{E.}~\bibnamefont{Gull}},
  \bibinfo{author}{\bibfnamefont{D.~R.} \bibnamefont{Reichman}},
  \bibnamefont{and} \bibinfo{author}{\bibfnamefont{A.~J.}
  \bibnamefont{Millis}}, \bibinfo{journal}{Phys. Rev. Lett.}
  \textbf{\bibinfo{volume}{112}}, \bibinfo{pages}{146802}
  (\bibinfo{year}{2014}),
  \urlprefix\url{http://link.aps.org/doi/10.1103/PhysRevLett.112.146802}.

\bibitem[{\citenamefont{Limot et~al.}(2005)\citenamefont{Limot, Kr\"oger,
  Berndt, Garcia-Lekue, and Hofer}}]{Limot_2005}
\bibinfo{author}{\bibfnamefont{L.}~\bibnamefont{Limot}},
  \bibinfo{author}{\bibfnamefont{J.}~\bibnamefont{Kr\"oger}},
  \bibinfo{author}{\bibfnamefont{R.}~\bibnamefont{Berndt}},
  \bibinfo{author}{\bibfnamefont{A.}~\bibnamefont{Garcia-Lekue}},
  \bibnamefont{and} \bibinfo{author}{\bibfnamefont{W.~A.} \bibnamefont{Hofer}},
  \bibinfo{journal}{Phys. Rev. Lett.} \textbf{\bibinfo{volume}{94}},
  \bibinfo{pages}{126102} (\bibinfo{year}{2005}),
  \urlprefix\url{http://link.aps.org/doi/10.1103/PhysRevLett.94.126102}.

\bibitem[{\citenamefont{Ternes et~al.}(2011)\citenamefont{Ternes, Gonz\'alez,
  Lutz, Hapala, Giessibl, Jel\'{\i}nek, and Heinrich}}]{Ternes_2011}
\bibinfo{author}{\bibfnamefont{M.}~\bibnamefont{Ternes}},
  \bibinfo{author}{\bibfnamefont{C.}~\bibnamefont{Gonz\'alez}},
  \bibinfo{author}{\bibfnamefont{C.~P.} \bibnamefont{Lutz}},
  \bibinfo{author}{\bibfnamefont{P.}~\bibnamefont{Hapala}},
  \bibinfo{author}{\bibfnamefont{F.~J.} \bibnamefont{Giessibl}},
  \bibinfo{author}{\bibfnamefont{P.}~\bibnamefont{Jel\'{\i}nek}},
  \bibnamefont{and} \bibinfo{author}{\bibfnamefont{A.~J.}
  \bibnamefont{Heinrich}}, \bibinfo{journal}{Phys. Rev. Lett.}
  \textbf{\bibinfo{volume}{106}}, \bibinfo{pages}{016802}
  (\bibinfo{year}{2011}),
  \urlprefix\url{http://link.aps.org/doi/10.1103/PhysRevLett.106.016802}.

\bibitem[{\citenamefont{N\'eel et~al.}(2009)\citenamefont{N\'eel, Kr\"oger, and
  Berndt}}]{Neel_2009}
\bibinfo{author}{\bibfnamefont{N.}~\bibnamefont{N\'eel}},
  \bibinfo{author}{\bibfnamefont{J.}~\bibnamefont{Kr\"oger}}, \bibnamefont{and}
  \bibinfo{author}{\bibfnamefont{R.}~\bibnamefont{Berndt}},
  \bibinfo{journal}{Phys. Rev. Lett.} \textbf{\bibinfo{volume}{102}},
  \bibinfo{pages}{086805} (\bibinfo{year}{2009}).

\bibitem[{\citenamefont{Tao et~al.}(2010)\citenamefont{Tao, Rungger, Sanvito,
  and Stepanyuk}}]{Tao_2010}
\bibinfo{author}{\bibfnamefont{K.}~\bibnamefont{Tao}},
  \bibinfo{author}{\bibfnamefont{I.}~\bibnamefont{Rungger}},
  \bibinfo{author}{\bibfnamefont{S.}~\bibnamefont{Sanvito}}, \bibnamefont{and}
  \bibinfo{author}{\bibfnamefont{V.~S.} \bibnamefont{Stepanyuk}},
  \bibinfo{journal}{Phys. Rev. B} \textbf{\bibinfo{volume}{82}},
  \bibinfo{pages}{085412} (\bibinfo{year}{2010}).

\bibitem[{\citenamefont{Frota and Oliveira}(1986)}]{Frota_1986}
\bibinfo{author}{\bibfnamefont{H.~O.} \bibnamefont{Frota}} \bibnamefont{and}
  \bibinfo{author}{\bibfnamefont{L.~N.} \bibnamefont{Oliveira}},
  \bibinfo{journal}{Phys. Rev. B} \textbf{\bibinfo{volume}{33}},
  \bibinfo{pages}{7871} (\bibinfo{year}{1986}),
  \urlprefix\url{http://link.aps.org/doi/10.1103/PhysRevB.33.7871}.

\bibitem[{\citenamefont{Perdew et~al.}(1996)\citenamefont{Perdew, Burke, and
  Ernzerhof}}]{Perdew1996}
\bibinfo{author}{\bibfnamefont{J.~P.} \bibnamefont{Perdew}},
  \bibinfo{author}{\bibfnamefont{K.}~\bibnamefont{Burke}}, \bibnamefont{and}
  \bibinfo{author}{\bibfnamefont{M.}~\bibnamefont{Ernzerhof}},
  \bibinfo{journal}{Phys.\ Rev.\ Lett.} \textbf{\bibinfo{volume}{77}},
  \bibinfo{pages}{3865} (\bibinfo{year}{1996}).

\bibitem[{\citenamefont{Kresse and Hafner}(1993{\natexlab{a}})}]{Kresse1993a}
\bibinfo{author}{\bibfnamefont{G.}~\bibnamefont{Kresse}} \bibnamefont{and}
  \bibinfo{author}{\bibfnamefont{J.}~\bibnamefont{Hafner}},
  \bibinfo{journal}{Phys. Rev. B} \textbf{\bibinfo{volume}{47}},
  \bibinfo{pages}{558} (\bibinfo{year}{1993}{\natexlab{a}}).

\bibitem[{\citenamefont{Kresse and Hafner}(1993{\natexlab{b}})}]{Kresse1993b}
\bibinfo{author}{\bibfnamefont{G.}~\bibnamefont{Kresse}} \bibnamefont{and}
  \bibinfo{author}{\bibfnamefont{J.}~\bibnamefont{Hafner}},
  \bibinfo{journal}{Phys. Rev. B} \textbf{\bibinfo{volume}{48}},
  \bibinfo{pages}{13115} (\bibinfo{year}{1993}{\natexlab{b}}).

\bibitem[{\citenamefont{Kresse and
  Furthm{\"u}ller}(1996{\natexlab{a}})}]{Kresse1996a}
\bibinfo{author}{\bibfnamefont{G.}~\bibnamefont{Kresse}} \bibnamefont{and}
  \bibinfo{author}{\bibfnamefont{J.}~\bibnamefont{Furthm{\"u}ller}},
  \bibinfo{journal}{Comput. Mater. Sci.} \textbf{\bibinfo{volume}{6}},
  \bibinfo{pages}{15} (\bibinfo{year}{1996}{\natexlab{a}}).

\bibitem[{\citenamefont{Kresse and
  Furthm{\"u}ller}(1996{\natexlab{b}})}]{Kresse1996b}
\bibinfo{author}{\bibfnamefont{G.}~\bibnamefont{Kresse}} \bibnamefont{and}
  \bibinfo{author}{\bibfnamefont{J.}~\bibnamefont{Furthm{\"u}ller}},
  \bibinfo{journal}{Phys. Rev. B} \textbf{\bibinfo{volume}{54}},
  \bibinfo{pages}{11169} (\bibinfo{year}{1996}{\natexlab{b}}).

\bibitem[{\citenamefont{Kresse and Joubert}(1999)}]{Kresse1999}
\bibinfo{author}{\bibfnamefont{G.}~\bibnamefont{Kresse}} \bibnamefont{and}
  \bibinfo{author}{\bibfnamefont{D.}~\bibnamefont{Joubert}},
  \bibinfo{journal}{Phys. Rev. B} \textbf{\bibinfo{volume}{59}},
  \bibinfo{pages}{1758} (\bibinfo{year}{1999}).

\bibitem[{\citenamefont{Hafner}(2008)}]{Hafner2008}
\bibinfo{author}{\bibfnamefont{J.}~\bibnamefont{Hafner}}, \bibinfo{journal}{J.\
  Comput.\ Chem.} \textbf{\bibinfo{volume}{29}}, \bibinfo{pages}{2044}
  (\bibinfo{year}{2008}).

\bibitem[{\citenamefont{Soler et~al.}(2002)\citenamefont{Soler, Artacho, Gale,
  Garc\'{\i}a, Junquera, Ordej\'on, and S\'anchez-Portal}}]{Soler2002}
\bibinfo{author}{\bibfnamefont{J.~M.} \bibnamefont{Soler}},
  \bibinfo{author}{\bibfnamefont{E.}~\bibnamefont{Artacho}},
  \bibinfo{author}{\bibfnamefont{J.~D.} \bibnamefont{Gale}},
  \bibinfo{author}{\bibfnamefont{A.}~\bibnamefont{Garc\'{\i}a}},
  \bibinfo{author}{\bibfnamefont{J.}~\bibnamefont{Junquera}},
  \bibinfo{author}{\bibfnamefont{P.}~\bibnamefont{Ordej\'on}},
  \bibnamefont{and}
  \bibinfo{author}{\bibfnamefont{D.}~\bibnamefont{S\'anchez-Portal}},
  \bibinfo{journal}{Journal of Physics: Condensed Matter}
  \textbf{\bibinfo{volume}{14}}, \bibinfo{pages}{2745} (\bibinfo{year}{2002}),
  \urlprefix\url{http://dx.doi.org/10.1088/0953-8984/14/11/302}.

\bibitem[{\citenamefont{Artacho et~al.}(2008)\citenamefont{Artacho, Anglada,
  Di\'eguez, Gale, Garc\'{\i}a, Junquera, Martin, OrdejÃ³n, Pruneda,
  SÃ¡nchez-Portal et~al.}}]{Artacho2008}
\bibinfo{author}{\bibfnamefont{E.}~\bibnamefont{Artacho}},
  \bibinfo{author}{\bibfnamefont{E.}~\bibnamefont{Anglada}},
  \bibinfo{author}{\bibfnamefont{O.}~\bibnamefont{Di\'eguez}},
  \bibinfo{author}{\bibfnamefont{J.~D.} \bibnamefont{Gale}},
  \bibinfo{author}{\bibfnamefont{A.}~\bibnamefont{Garc\'{\i}a}},
  \bibinfo{author}{\bibfnamefont{J.}~\bibnamefont{Junquera}},
  \bibinfo{author}{\bibfnamefont{R.~M.} \bibnamefont{Martin}},
  \bibinfo{author}{\bibfnamefont{P.}~\bibnamefont{OrdejÃ³n}},
  \bibinfo{author}{\bibfnamefont{J.~M.} \bibnamefont{Pruneda}},
  \bibinfo{author}{\bibfnamefont{D.}~\bibnamefont{SÃ¡nchez-Portal}},
  \bibnamefont{et~al.}, \bibinfo{journal}{J. Phys.: Condens. Matter}
  \textbf{\bibinfo{volume}{20}}, \bibinfo{pages}{064208}
  (\bibinfo{year}{2008}),
  \urlprefix\url{http://dx.doi.org/10.1088/0953-8984/20/6/064208}.

\bibitem[{\citenamefont{Brandbyge et~al.}(2002)\citenamefont{Brandbyge, Mozos,
  Ordejon, Taylor, and Stokbro}}]{Brandbyge2002}
\bibinfo{author}{\bibfnamefont{M.}~\bibnamefont{Brandbyge}},
  \bibinfo{author}{\bibfnamefont{J.}~\bibnamefont{Mozos}},
  \bibinfo{author}{\bibfnamefont{P.}~\bibnamefont{Ordejon}},
  \bibinfo{author}{\bibfnamefont{J.}~\bibnamefont{Taylor}}, \bibnamefont{and}
  \bibinfo{author}{\bibfnamefont{K.}~\bibnamefont{Stokbro}},
  \bibinfo{journal}{Phys. Rev. B} \textbf{\bibinfo{volume}{65}},
  \bibinfo{pages}{165401} (\bibinfo{year}{2002}).

\bibitem[{\citenamefont{Abufager et~al.}(2015)\citenamefont{Abufager, Robles,
  and Lorente}}]{Abufager2015}
\bibinfo{author}{\bibfnamefont{P.}~\bibnamefont{Abufager}},
  \bibinfo{author}{\bibfnamefont{R.}~\bibnamefont{Robles}}, \bibnamefont{and}
  \bibinfo{author}{\bibfnamefont{N.}~\bibnamefont{Lorente}},
  \bibinfo{journal}{J. Phys. Chem. C} \textbf{\bibinfo{volume}{119}},
  \bibinfo{pages}{12119} (\bibinfo{year}{2015}),
  \urlprefix\url{http://pubs.acs.org/doi/abs/10.1021/acs.jpcc.5b01839}.

\bibitem[{\citenamefont{Polok et~al.}(2011)\citenamefont{Polok, Fedorov,
  Bagrets, Zahn, and Mertig}}]{Polok_2011}
\bibinfo{author}{\bibfnamefont{M.}~\bibnamefont{Polok}},
  \bibinfo{author}{\bibfnamefont{D.~V.} \bibnamefont{Fedorov}},
  \bibinfo{author}{\bibfnamefont{A.}~\bibnamefont{Bagrets}},
  \bibinfo{author}{\bibfnamefont{P.}~\bibnamefont{Zahn}}, \bibnamefont{and}
  \bibinfo{author}{\bibfnamefont{I.}~\bibnamefont{Mertig}},
  \bibinfo{journal}{Phys. Rev. B} \textbf{\bibinfo{volume}{83}},
  \bibinfo{pages}{245426} (\bibinfo{year}{2011}),
  \urlprefix\url{http://link.aps.org/doi/10.1103/PhysRevB.83.245426}.

\bibitem[{\citenamefont{Surer et~al.}(2012)\citenamefont{Surer, Troyer, Werner,
  Wehling, L\"auchli, Wilhelm, and Lichtenstein}}]{Surer_2012}
\bibinfo{author}{\bibfnamefont{B.}~\bibnamefont{Surer}},
  \bibinfo{author}{\bibfnamefont{M.}~\bibnamefont{Troyer}},
  \bibinfo{author}{\bibfnamefont{P.}~\bibnamefont{Werner}},
  \bibinfo{author}{\bibfnamefont{T.~O.} \bibnamefont{Wehling}},
  \bibinfo{author}{\bibfnamefont{A.~M.} \bibnamefont{L\"auchli}},
  \bibinfo{author}{\bibfnamefont{A.}~\bibnamefont{Wilhelm}}, \bibnamefont{and}
  \bibinfo{author}{\bibfnamefont{A.~I.} \bibnamefont{Lichtenstein}},
  \bibinfo{journal}{Phys. Rev. B} \textbf{\bibinfo{volume}{85}},
  \bibinfo{pages}{085114} (\bibinfo{year}{2012}),
  \urlprefix\url{http://link.aps.org/doi/10.1103/PhysRevB.85.085114}.

\bibitem[{\citenamefont{Baruselli et~al.}(2015)\citenamefont{Baruselli,
  Requist, Smogunov, Fabrizio, and Tosatti}}]{Baruselli_2015}
\bibinfo{author}{\bibfnamefont{P.~P.} \bibnamefont{Baruselli}},
  \bibinfo{author}{\bibfnamefont{R.}~\bibnamefont{Requist}},
  \bibinfo{author}{\bibfnamefont{A.}~\bibnamefont{Smogunov}},
  \bibinfo{author}{\bibfnamefont{M.}~\bibnamefont{Fabrizio}}, \bibnamefont{and}
  \bibinfo{author}{\bibfnamefont{E.}~\bibnamefont{Tosatti}},
  \bibinfo{journal}{Phys. Rev. B} \textbf{\bibinfo{volume}{92}},
  \bibinfo{pages}{045119} (\bibinfo{year}{2015}),
  \urlprefix\url{http://link.aps.org/doi/10.1103/PhysRevB.92.045119}.

\bibitem[{\citenamefont{Jacob}(2015)}]{Jacob_2015}
\bibinfo{author}{\bibfnamefont{D.}~\bibnamefont{Jacob}},
  \bibinfo{journal}{Journal of Physics: Condensed Matter}
  \textbf{\bibinfo{volume}{27}}, \bibinfo{pages}{245606}
  (\bibinfo{year}{2015}),
  \urlprefix\url{http://stacks.iop.org/0953-8984/27/i=24/a=245606}.

\bibitem[{\citenamefont{Frank and Jacob}(2015)}]{Frank_2015}
\bibinfo{author}{\bibfnamefont{S.}~\bibnamefont{Frank}} \bibnamefont{and}
  \bibinfo{author}{\bibfnamefont{D.}~\bibnamefont{Jacob}},
  \bibinfo{journal}{Phys. Rev. B} \textbf{\bibinfo{volume}{92}},
  \bibinfo{pages}{235127} (\bibinfo{year}{2015}),
  \urlprefix\url{http://link.aps.org/doi/10.1103/PhysRevB.92.235127}.

\bibitem[{\citenamefont{Dang et~al.}(2016)\citenamefont{Dang, dos Santos~Dias,
  Liebsch, and Lounis}}]{Lounis_2016}
\bibinfo{author}{\bibfnamefont{H.~T.} \bibnamefont{Dang}},
  \bibinfo{author}{\bibfnamefont{M.}~\bibnamefont{dos Santos~Dias}},
  \bibinfo{author}{\bibfnamefont{A.}~\bibnamefont{Liebsch}}, \bibnamefont{and}
  \bibinfo{author}{\bibfnamefont{S.}~\bibnamefont{Lounis}},
  \bibinfo{journal}{Phys. Rev. B} \textbf{\bibinfo{volume}{93}},
  \bibinfo{pages}{115123} (\bibinfo{year}{2016}),
  \urlprefix\url{http://link.aps.org/doi/10.1103/PhysRevB.93.115123}.

\bibitem[{\citenamefont{Meir and Wingreen}(1992)}]{Meir}
\bibinfo{author}{\bibfnamefont{Y.}~\bibnamefont{Meir}} \bibnamefont{and}
  \bibinfo{author}{\bibfnamefont{N.~S.} \bibnamefont{Wingreen}},
  \bibinfo{journal}{Phys. Rev. Lett.} \textbf{\bibinfo{volume}{68}},
  \bibinfo{pages}{2512} (\bibinfo{year}{1992}),
  \urlprefix\url{http://link.aps.org/doi/10.1103/PhysRevLett.68.2512}.

\bibitem[{\citenamefont{Koryt\'ar and Lorente}(2011)}]{Korytar_2011}
\bibinfo{author}{\bibfnamefont{R.}~\bibnamefont{Koryt\'ar}} \bibnamefont{and}
  \bibinfo{author}{\bibfnamefont{N.}~\bibnamefont{Lorente}},
  \bibinfo{journal}{Journal of Physics: Condensed Matter}
  \textbf{\bibinfo{volume}{23}}, \bibinfo{pages}{355009}
  (\bibinfo{year}{2011}),
  \urlprefix\url{http://stacks.iop.org/0953-8984/23/i=35/a=355009}.

\end{thebibliography}
\bibliographystyle{apsrev}

\end{document}